\documentclass[%
 reprint,
 amsmath,amssymb,
prl,
]{revtex4-1}

\usepackage{graphicx}% Include Figure files
\usepackage{dcolumn}% Align table columns on decimal point
\usepackage{bm}% bold math
\begin{document}

%\preprint{APS/123-QED}

\title{{\it In situ} high resolution real-time  quantum efficiency imaging for photocathodes}% Force line breaks with \\
%%%%
\thanks{Work supported by China National Key Scientific Instrument and Equipment Development Project (grant no. 2011YQ130018), National Natural Science Foundation of China (grant nos 11475159, 11505173, 11576254 and 11605190)}%

\author{Dai Wu}
\email{wudai04@gmail.com}
 \author{Dexin Xiao}
 \author{Jianxin Wang}
 \author{Qing Pan}
 \author{Xing Luo}
 \author{Kui Zhou}
 \author{Chenglong Lao}
 \author{Xiangkun Li}
 \author{Sifen Lin}
 \author{Peng Li}
 \author{Hao Zhang}
 \author{Longgang Yan}
 \author{Hanbin Wang}
 \author{Xingfan Yang}
  \author{Ming Li}%
 \email{liming@caep.cn}

\affiliation{%
 Institute of Applied Electronics, China Academy of Engineering Physics (CAEP/IAE),
Mianyang, 621900, P.R.China\\
}%


%%%%%%%%%%%
\author{Chuanxiang Tang}
\author{Dan Wang}
\affiliation{
 Department of Engineering Physics, Tsinghua University, Beijing, 100084, P.R.China\\
}%
\affiliation{
Key Laboratory of Particle and Radiation Imaging (Tsinghua University), Ministry of Education,
Beijing, 100084, P. R. China
}%
\author{Qilong Wang}
\author{Zhiyang Qi}
\author{Jian Zhang}
\affiliation{%
 School of Electronic Science and Engineering, Southeast University, Nanjing,211189, P.R.China\\
}%
%%%%%%%%%%%%%%%%%%%%%%%%%第二个删除部分的结尾%%%%%%%%%%%%%%


\date{\today}% It is always \today, today,

\begin{abstract}
Aspects of the preparation process and performance degradation are two major problems of photocathodes.
 The lack of a means for dynamic quantum efficiency measurements results in the inability to observe the inhomogeneity of the cathode surface by fine structural analysis and in real time.
Here we present a simple and scalable technique for {\it in situ} real-time quantum efficiency diagnosis.
An incoherent light source provides uniform illumination on the cathode
surface, and solenoid magnets are used as lens for focusing and imaging the emitted electron beam on a downstream scintillator screen, which converts the quantum efficiency information into fluorescence intensity distribution.
The microscopic discontinuity and the dynamic changes of the quantum efficiency of a gallium arsenide photocathode are observed at a resolution of a few microns.
An unexpected uneven decrease of the quantum efficiency is also recorded.
The work demonstrates a new observation method for photoemission materials research.

\begin{description}
% Secondary publications and information retrieval purposes.
\item[PACS numbers]
85.60.Ha, 29.20.Ej
% You may use the \texttt{description} environment to structure your abstract;
% use the optional argument of the \verb+\item+ command to give the category of each item.
\end{description}
\end{abstract}

\pacs{Valid PACS appear here}% PACS, the Physics and Astronomy
\maketitle  %%%%%%%%%第一个

%\tableofcontents

%\section{\label{sec:level1}Introduction}
Over the past several decades, photocathodes have played various important roles in many fields, such as image intensifier \cite{biberman2013photoelectronic}, ultra-fast streak camera \cite{kassier2010compact}, and photomultiplier tube \cite{engstrom1980photomultiplier}, etc.
Especially in recent years, photocathodes have become the most important high-brightness electron sources to drive X-ray free electron lasers or Compton back-scattering $\gamma$-ray sources \cite{dowell1993first,emma2010first,dowell2010cathode,Dowell2009,PhysRevSTAB.11.030703,ding2009measurements,Zheng201698}, which have promoted the development of optical sciences, biological sciences, organic chemistry and many other cutting-edge sciences \cite{chapman2009x,rohringer2012atomic,hirata2014determination,suga2015native}.

Although photocathodes have been widely used, there are still some questions surrounding them, among which the microscopic dynamic process
of their performance, which mostly refers to the quantum efficiency (QE),
%of the incident photon to converted electron ratio,
is the utmost importance.
For example, the relationship between the photocathode QE degradation and the environment, such as temperature or ion back-bombardment, has been widely studied \cite{fischer1994thermal,Aulenbacher1997498,lewellen2002ion,qiang2010particle,Zhang2011a,grames2011charge}, but the details are mostly hypothetical or based on simulation. Also, field emission defects and the non-uniformity of photocathode's surface will reduce the quality of the high-brightness electron beam and the X-ray or $\gamma$-ray sources driven by it, making laser transverse shaping fails to some extent, but it is complicated to measure the QE uniformity in the small range of the laser irradiation spot \cite{pandey2009field,shao2016situ}. In addition, the dynamic effect of each component on the cathode performance during the preparation or activation of the semiconductor photocathode is typically measured only by laser irradiation at a single spot.
All these issues arise from the lack of a means for {\it in situ} high-resolution real-time QE observation. Earlier studies used a laser focused to a small spot to raster scan the QE of the cathode surface. This method is called 'QE mapping', the resolution of which could reach $<100$ $\mu$m  \cite{grames2011charge,Gubeli2001554,sinclair2007development,cultrera2011thermal,PhysRevSTAB.15.090703,riddick2013photocathode,filippetto2015cesium},
but the scanning time would last for minutes to hours, depending on the resolution.
Photoemission electron microscopy (PEEM) could offer the best resolution ($\sim 10$ nm) of QE distribution  \cite{PEEM1999,PhysRevLett.110.076802}. However, this technique requires the cathode to be moved into its system, thus the photocathodes cannot be observed in real-time or {\it in situ}.
%In particular, it is not suitable for numerous photocathodes that cannot be removed from their vacuum environment.
%%

In this paper, we present a simple {\it in situ} real-time QE imaging of a high-voltage photocathode DC gun, as well as some of the cathode dynamic changes observed during its use.

This study was conducted at the Tera-Hertz free electron laser (THz FEL) facility \cite{wudaiSAP2014} at  the China Academy of Engineering Physics (CAEP). The electron beam line downstream the high-voltage DC gun was used as the imaging system, which is shown in Fig. 1(a). In order to avoid the coupling of the transverse distribution of the coherent light source, a lamp, as an incoherent light source, was placed far enough from the photocathode, to provide a uniform illumination. Although QE is a function of the wavelength, a uniform illumination ensures that the transverse density distribution of the emitted electrons is proportional to the relative QE distribution at any corresponding wavelength. The emitted electron beam was then accelerated by the electric field inside the high-voltage gun. The voltage can reach 320 kV, which makes the field gradient on the cathode surface $E_0=$4.14 MV/m, as calculated by the SUPERFISH code \cite{poissonsuperfish}.
In this work, four pieces of gallium arsenide (GaAs) photocathodes were tested.
%These cathodes were activated by CS/O \cite{vergara1992adsorption}, with a work function of 1.4 eV.
All the cathodes were flat, 18 cm in diameter and covered by a tantalum cap.
%
%The cathodes could be loaded and unloaded in the gun with a 'load-lock' system.
%
% Although the electric field is perpendicular to the cathode surface, the electrons have  their intrinsic transverse momentum, resulting in an enlarged virtual image downstream the cathode.
 The average emission current was about a pico to a nano Ampere, thus the space charge effect can be neglected, and it was a linear electron optical system.
Two solenoids were placed downstream to provide two different images, one is an enlarged
 image with better resolution to give some specific local details, and the other image is smaller so as to observe the whole QE distribution over the cathode. The images are generated at the YAG phosphor screens and captured by the CCDs.
 %, which convert the electron density distribution into fluorescence intensity distribution captured by the CCDs.
 %The angle between the screen and the beamline is 45$^\circ$, to ensure that the fluorescence image captured by the charge-coupled device (CCD) is in the direction perpendicular to the beamline and is thus not distorted.
%
A laser scanning system focuses the laser to a specific point and measures the QE at that point,
%with the fast-current transformer (FCT),
making the absolute QE distribution of the entire cathode available.

\begin{figure}[!htp] % use float package if you want it here
  \centering
  \includegraphics*[width=220pt]{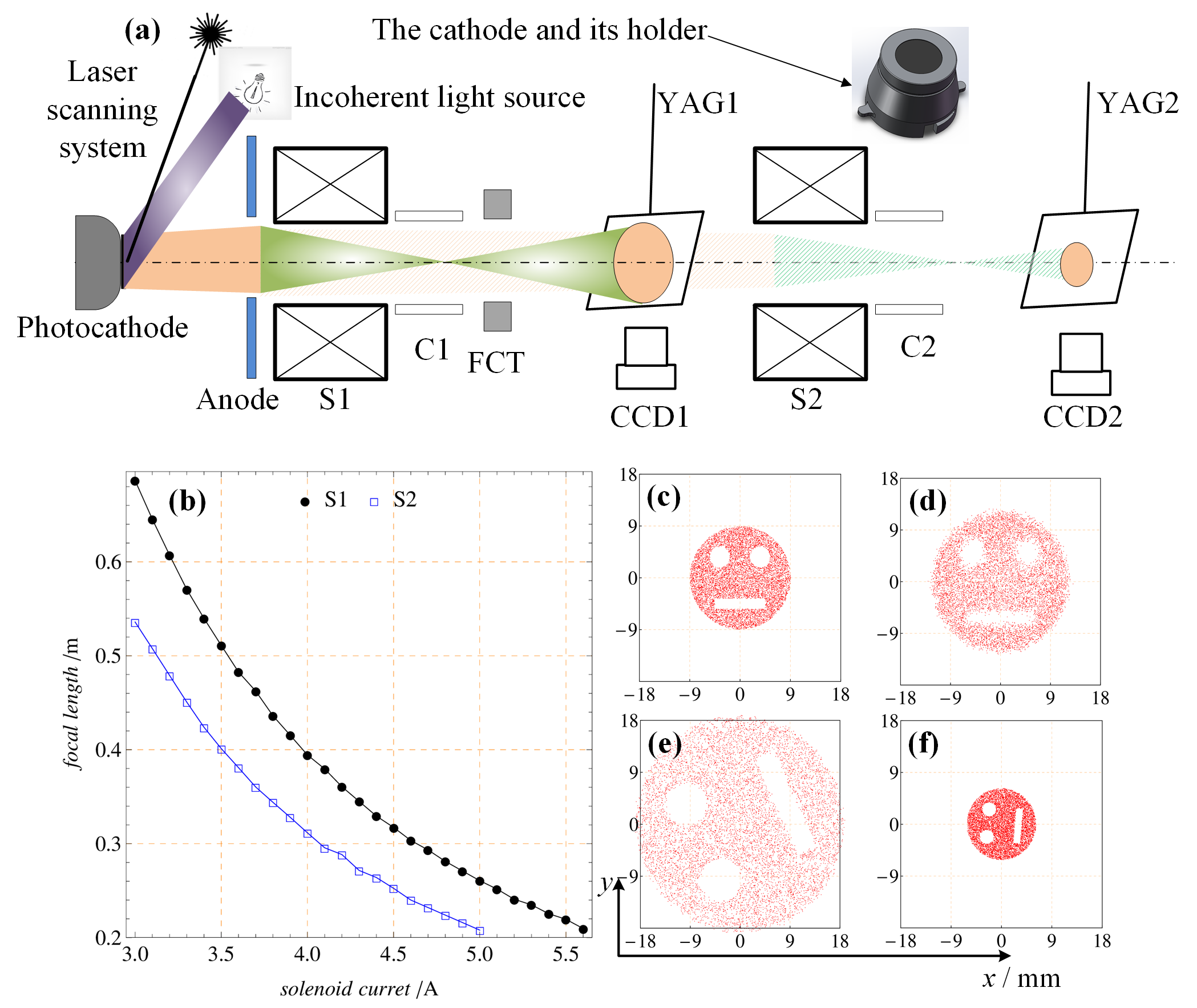}
  \caption{The photocathode QE imaging system at CAEP. (a) The beamline layout (Not to scale): S, Solenoid magnet to focus the electron beam; C, magnetic Corrector to change the direction of electron beam; FCT, Fast Current Transformer to measure the beam bunch charge;
  YAG, Yttrium Aluminum Garnet phosphor screen doped with cerium; CCD, Charge-Coupled Device for image detection. (b) The equivalent focal length of the solenoids versus excitation current. (c)-(f) ASTRA simulation results for electron distribution on the cathode (with an initial 'evil face' pattern), in the gun exit, at YAG1 and at YAG2.}
  \label{Fig:Fig1_all_20161213}
\end{figure}

%The cathode surface was set at $z$=0, and S1, S2, YAG1, YAG2 were placed at $z=$0.24 m, 1.4 m, 1.05 m and 1.9 m, respectively.
%For this linear electron optical system, the convex lens imaging formula can be used to estimate the appropriate focal length.
The equivalent focal length of the solenoids versus the excitation current, as simulated by the particle tracking code ASTRA \cite{Floettmann2011}, is shown in Fig. 1(b).
%To capture an enlarged image at YAG1, the excitation current of S1 should be more than 5.2 A. For the imaging at YAG2, the current of S1 and S2 should both be in the range from 3.8 A to 4.8 A.
%
The beam was rotated when passing through the solenoid.
An 'evil face' pattern was used to simulate the magnification and rotation.
In the simulation, macro-particles were generated by the code ASTRA with a uniform distribution and the same size as the photocathode, as shown in Fig. 1(c).
Fig. 1(d) shows the electron distribution outside the gun with no solenoid magnet, where the pattern turned to be larger and not clear due to the initial transverse momentum. The enlarged pattern, with the excitation current of S1 at 5.7 A, is shown in Fig. 1(e). In this case, the image magnification was about 2.1, and the rotation was about 116$^\circ$ counterclockwise.
When the current of S1 and S2 were 4.4 A and 4.0 A, respectively, a smaller image could be captured at YAG2, as shown in Fig. 1(f), where the magnification was about 0.6, and the rotation was 86$^\circ$ counterclockwise.
%Both the Fig. 1(e) and 1(f)  images were sharp enough to observe the details of the electron  distribution.

%Since the gun voltage was as high as 320 kV, the relative energy spread at the anode was only about 1.7\%, thus the system can be considered to be non-dispersive.
The transverse spatial resolution $R$ is mainly determined by the beamline optical magnification and the CCD resolution. The beamline optical magnification was limited by three factors, which are the vacuum pipe size ( 60 mm in diameter), the YAG screen size (25 mm wide and 18 mm high) and the maximum current of the power sources of S1 and S2 ( 6 A and 4 A, respectively).
For all these factors the magnification was found to be less than 3.The CCD is a model 'FL3-GE-13S2M-C' PointGrey camera \cite{pointgrey2013}, equipped with a Nikon 'AF Nikkor 50mm f/1.8D' lens. The best resolution of this CCD is about 30 $\mu$m/pixel, making the best resolution of the whole QE imaging system be about 10 $\mu$m/pixel. The dynamic range of the camera is 59.14 dB, which indicates that the imaging method can theoretically resolve QE differences as small as 10$^{-6}$.

\begin{figure}[!htp] %
  \centering
  \includegraphics*[width=240pt]{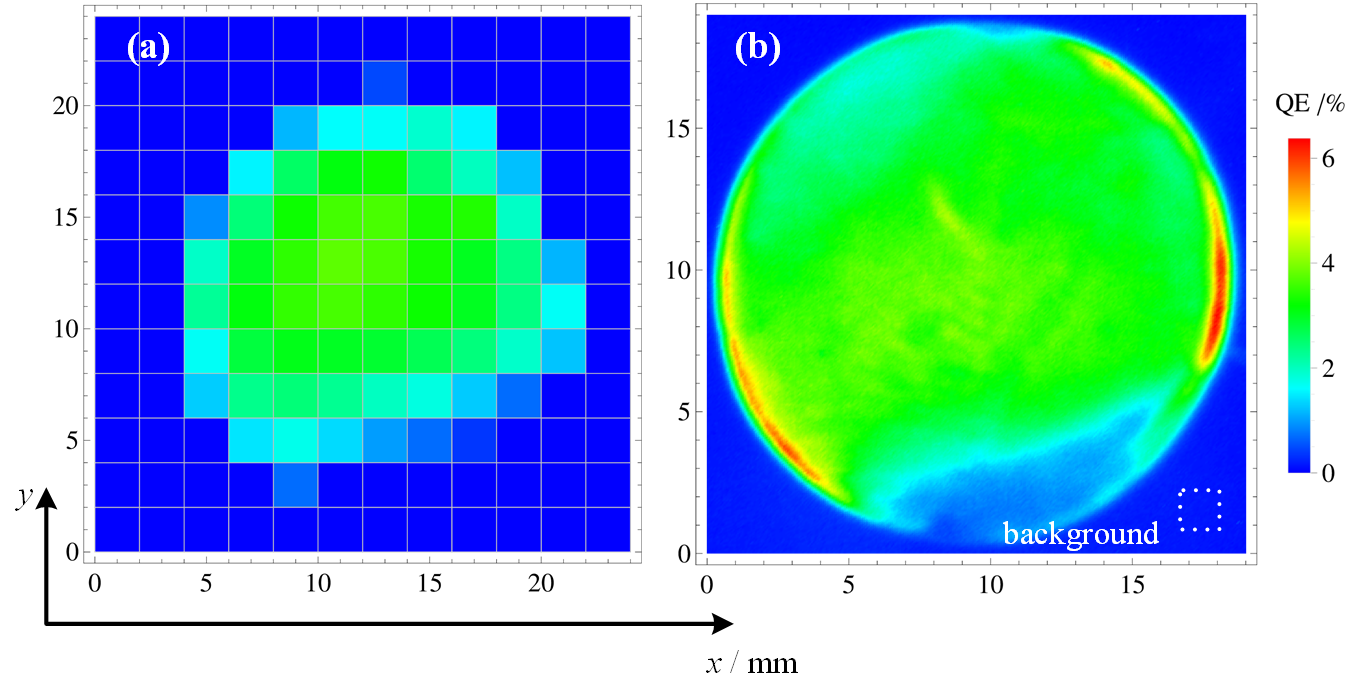}
  \caption{GaAs photocathode QE distribution. (a) QE mapping measured by the laser scanning across the cathode surface. (b) The QE imaging captured at YAG2.
  %when activation current of S1 and S2 was 4.1 A and 4.0 A, respectively.
  }
  \label{Fig:scan_vs_image_20161214}
\end{figure}

The results obtained by the two methods for measuring QE distribution of cathode \#1 are shown in Fig. 2.  For the former, the photo-injector drive laser was focused into a small spot to scan the QE with a step size of about 1.81 mm. The laser had an approximately round Gaussian transverse distribution with a root mean square (RMS) size of $\sigma_{r}=$0.64 mm. The space charge limit (SCL) was calculated as \cite{rao2014engineering}: ${Q_{{\rm{SCL}}}} = 2\pi {\varepsilon _0}{E_0}\sigma _r^2=94$ pC,%
where $\varepsilon_0$ is the dielectric constant in the vacuum. It means the emitted bunch charge would be proportional to the QE, as long as it is less than 94 pC. The laser power could be continuously adjusted in the range from 0 to 8 W at the wavelength of 532 nm. %
And the duty factor was adjustable from 5$\times 10^{-9}$ to 1.
%The repetition of the laser was 54.167 MHz and the RMS length was 5 ps.
%The laser was operated mostly in macro-pulse mode, the macro-pulse length can be adjusted from 5 ns to 1 ms, and the macro-repetition rate from 1 Hz to 1 kHz, making the duty factor adjustable from 5$\times 10^{-9}$ to 1.
The raster scanning QE distribution is shown in Fig. 2(a). The largest single bunch charge is 50.2 pC, less than 94 pC.
%Accordingly, the bunch charge distribution represents the QE distribution.

For comparison, the pseudo-color picture of the QE distribution captured at YAG2 is shown in Fig. 2(b). As the saturation was not reached, the brightness of the phosphor screen was proportional to the charge density, which was also proportional to the QE distribution.
The absolute value of the QE was given by a single spot of the scanning result in a relatively uniform region. The rotation of the image has been restored. Also, by cathode size calibration, the distance between two pixels represented about 82 $\mu$m on the cathode surface.
The image was sharper,
%than the one generated by laser raster scanning,
and most importantly it took much less time.
 The captured bitmap (BMP) images have only 256 levels of grayscale, so the dynamic range of this system was about 24 dB. The selected background within the dotted white box has an averaged 3.95 level of grayscale, making the signal-to-noise ratio 65:1.
 %While the laser power and electron bunch charge measurement error was about 2\%, the dynamic range of the raster scanning QE map was only 17 dB.

Neither of the two methods can obtain accurate results at the edge of the cathode. For the scanning method, the laser will be cut at the edge, thus the QE map is a little larger than the cathode in size. Meanwhile, for the imaging method, the lamp light is intensified by the tantalum cap, so the QE will be greater than the actual value at the edge. Both methods show that the lower right corner of the cathode has lower QE. These non-uniformities are due to the nonuniform growth of the Cs-O active layer on the GaAs cathode surface. The optimum number of atoms in the active layer is Cs:O $\approx$2:1 \cite{Clark1975}, but in the cathode preparation chamber, the two Cs sources are placed at both side of the cathode, one of which evaporates less than the other.
%

%As the QE image was captured {\it in situ},
The dynamic process of the QE degradation caused by ion bombardment was observed  {\it in situ} on the surface of cathode \#1, as shown in Fig. 3.  The electron gun was working in the continuous wave (CW) mode for about 2 hours to make the effect of the ion bombardment significant. The average current of the cathode emission was 0.5 mA.
%was achieved by adjusting the Glan polarizing prism.
The laser struck near the center of the cathode. The average laser power was under 50 mW to make sure that the temperature at the laser spot on the cathode did not change greatly (less than 1 $^\circ$C, simulated by Ansys code).
%
%When the current was emitting from the cathode, the vacuum inside the gun reduced from 3$\times 10^{-9}$ Pa to about 2$\times 10^{-8}$ Pa.
The QE of the photocathode should be exponentially decreased if only the influence of vacuum was considered. Accordingly, the ratio of the original QE image (Fig. 1(b)) to the new QE image after this CW operation should be flat. But in reality it shows obvious differences, as shown in the red dashed circle and the blue dashed ellipse of Fig. 3(a).

\begin{figure}[!htp] %
  \centering
  \includegraphics*[width=240pt]{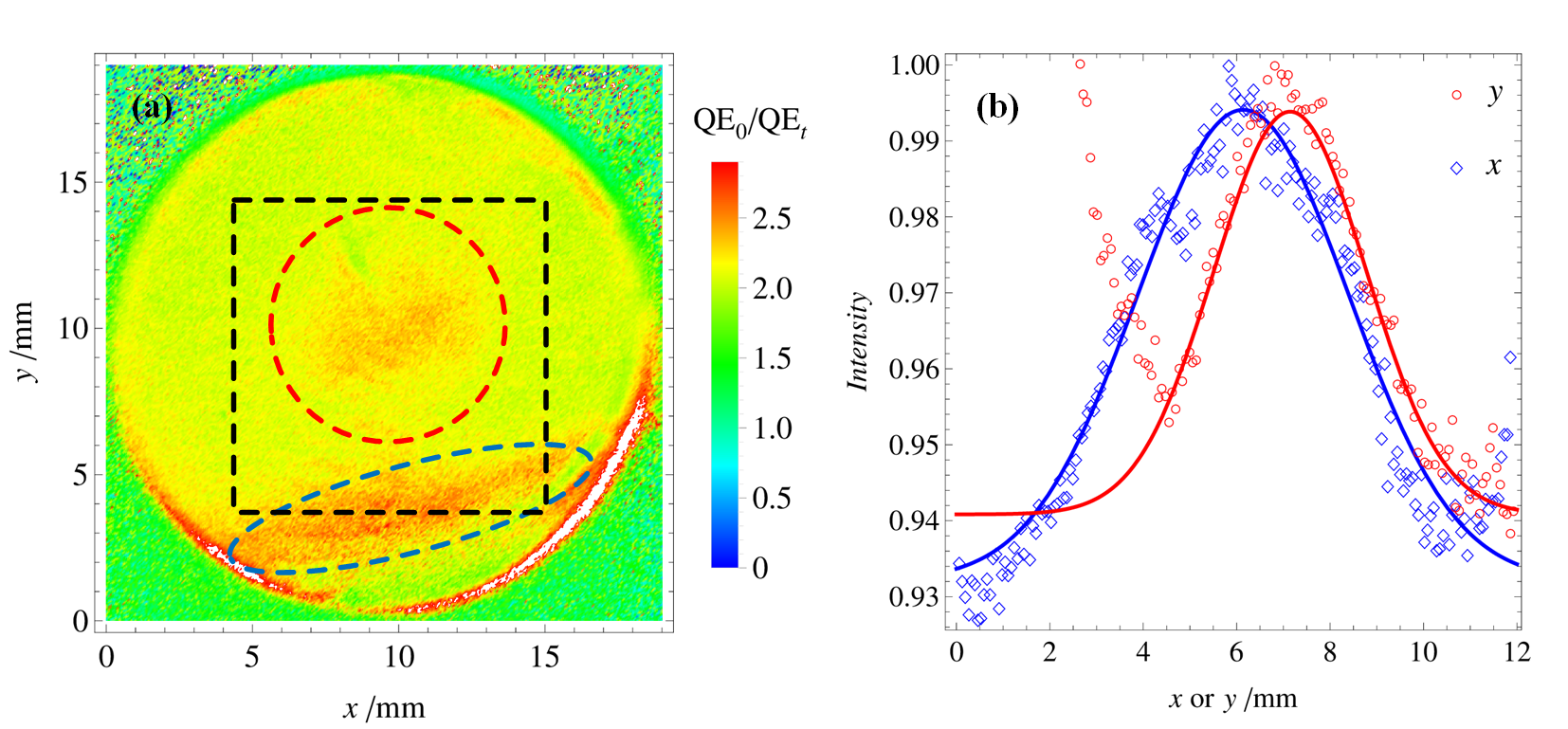}
  \caption{The GaAs photocathode ion back-bombardment distribution measurement with QE imaging. (a) The ratio of the original QE image to the new QE image after a period of CW operation on cathode \#1. The red dashed circle represents the image formed by ion back-bombardment. The blue dashed ellipse indicates that the non-uniform region of the active layer also has a faster rate of QE degradation. (b) The distribution projections in the $x$ and $y$ directions inside the black dashed square of (a).  }
  \label{Fig:ion_distribution_all_20161216}
\end{figure}

Since the cathode is flat and the electric field of the electron gun is radially symmetrical, the ions, produced by ionization in the vacuum system and accelerated by the electric field, would strike around the center, making the QE of that area degrade faster. The rate of decline, to some extent, represents the density of the ions. Therefore, the distribution within the red dashed circle represents the ion back-bombardment distribution and its effect on QE. The distribution projections in the $x$ and $y$ directions inside the black dashed square are shown in Fig. 3(b), which can be fitted as a Gaussian distribution.
%, and has an RMS of $\sigma_x=2.26$ mm and $\sigma_y=1.63$ mm, respectively.
The results were similar to those obtained by the Monte Carlo simulation in the RF gun \cite{qiang2010particle}.

It could also be observed that the QE near the lower edge of the  photocathode decreased more rapidly than it did in other areas, as shown in the blue dashed ellipse in Fig. 3(a). This phenomenon is an evidence of different dark life-time related to different activation layers  across the GaAs photocathode, which suggests a way of improving the preparation process to obtain uniform QE on the cathode surface. However, the areas above and below the ellipse had almost the same QE descending speed, which is worthy of further study.

\begin{figure}[!htp]
  \centering
  \includegraphics*[width=240pt]{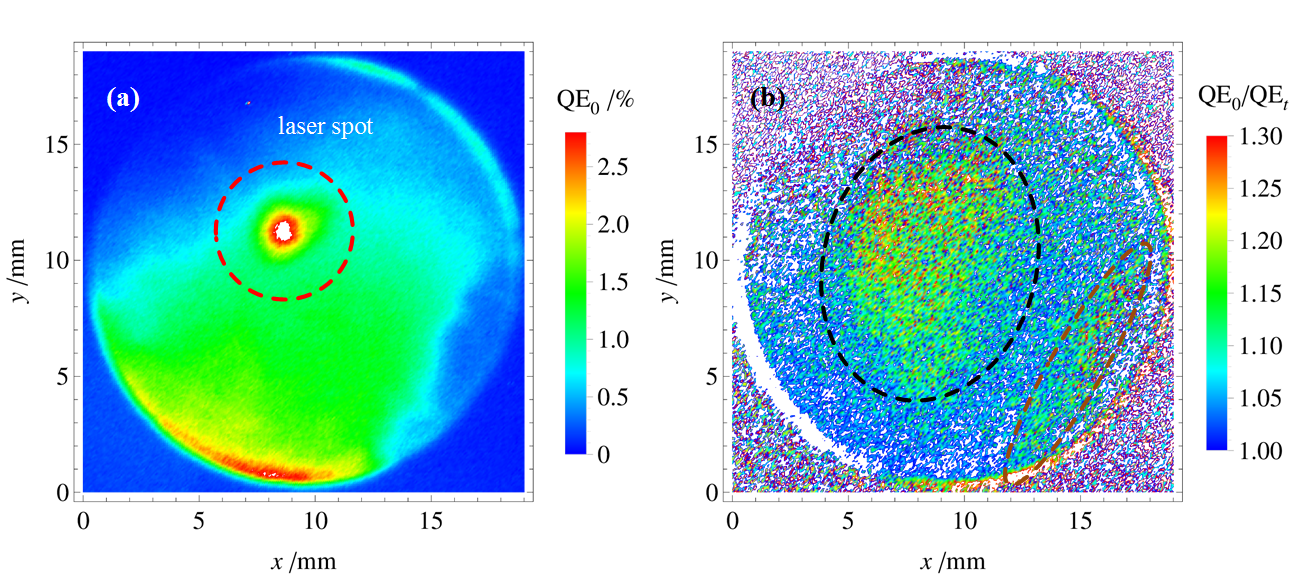}
  \caption{QE degradation due to temperature difference. (a) The original QE image on cathode \#2. The red dashed circle represents the location of the laser irradiation. (b)
  The ratio of the original QE image to the new QE image after 1 W average power laser irradiation with the high voltage turned off. The black dashed ellipse represents QE difference caused by laser heating. The red ellipse indicates a faster rate of QE degradation at the edge of the non-uniform area again.}
  \label{Fig:QE_temp_all_20161215}
\end{figure}

The QE degradation difference due to temperature difference was also studied.
%with the imaging system.
An image of photocathode \#2 with more non-uniform QE distribution was captured
%, using an activation current of 4.4 A and 4.0 A for S1 and S2, respectively.
By adjusting the laser to a very low average power, the emitted electron bunch generated by the laser irradiation would have a very small charge density. Therefore, the space charge effect could be ignored, and the laser distribution could be imaged together with the QE imaging, when the laser irradiates a relatively uniform region on the cathode. This would be an {\it in situ} laser transverse distribution imaging method. The laser RMS size of $\sigma_r=0.64$ mm was measured in this way and the result was in good agreement with the one measured by the laser profile CCD.
The laser spot location was within the red dashed circle in Fig. 4(a).
%to indicate the irradiation location.

In order to avoid ion back-bombardment-induced QE degradation, the high-voltage of the DC gun was turned off, and no electrons were emitted from the cathode surface to generate ions. The laser was then operated in continuous mode to heat the cathode surface with a 1 W average power for 35 minutes. When the laser was striking the cathode, the vacuum inside the gun remained at about 3$\times 10^{-9}$ Pa. The ratio of the original QE image to the new one after this CW operation was not flat ether. With the laser irradiation on the center, a large area showed a QE decrease of about 10\% $\sim$30\%, as shown in the black dashed ellipse in Fig. 4(b).

\begin{figure}[!htp] %
  \centering
  \includegraphics*[width=240pt]{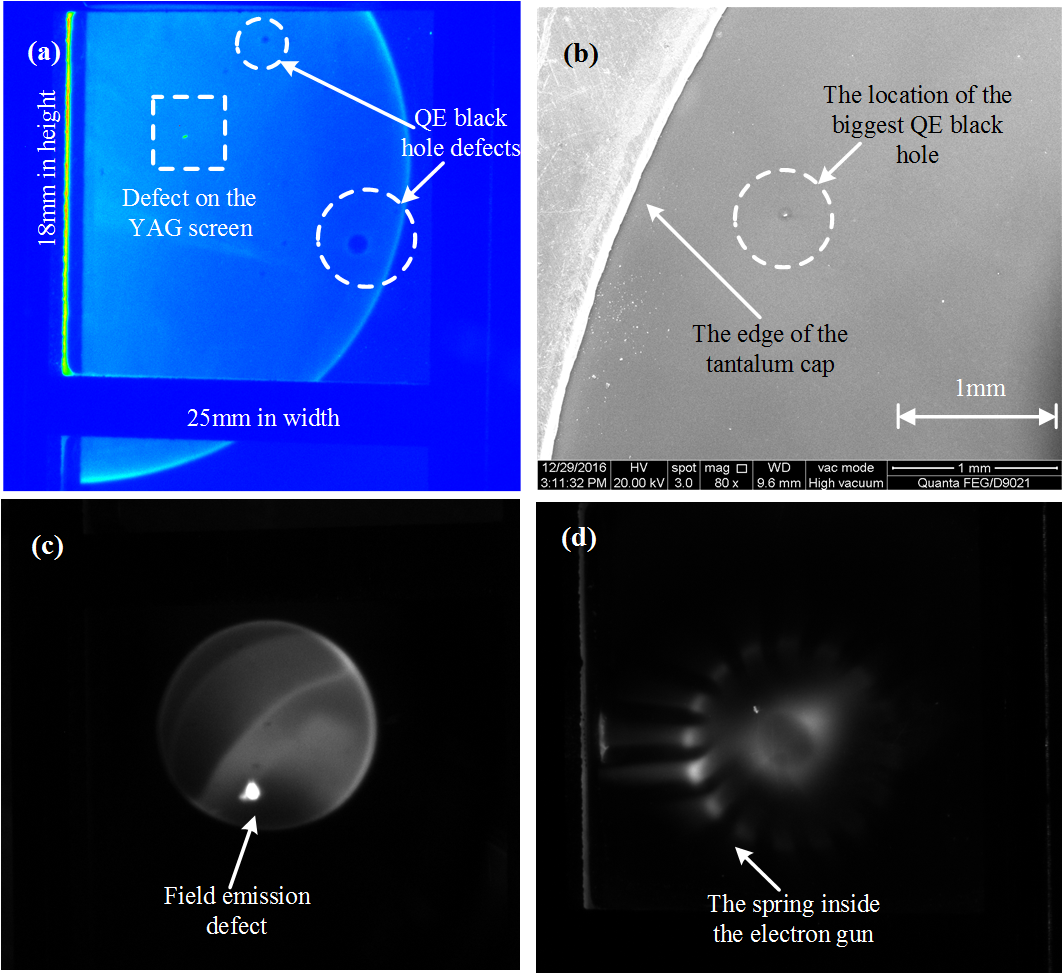}
  \caption{The imaging of QE defects and high-voltage breakdown. (a) The enlarged image of the QE distribution of cathode \#3 at YAG1.
  There were two YAG crystals mounted next to each other. The white dashed square represents a defect on the YAG crystal itself. The white dashed circles show the location of some QE black hole defects. (b)The imaging around the location of the biggest defect captured by scanning electron microscopy (SEM). (c) A typical QE imaging with a field emission defect on the cathode \#4. (d) An imaging of high-voltage breakdown when the cathode is unloaded. }
  \label{Fig:defects_all_20161215}
\end{figure}

Some lower QE defects like a kind of 'black hole' were observed on the enlarged QE image, as shown in Fig. 5(a). The activation current of S1 is 5.7 A, corresponding to an image magnification factor of about 2.1. The resolution of this image is about 36 $\mu$m/pixel, making the resolution on the cathode surface about 17 $\mu$m/pixel.
The biggest 'black hole' is about 391 $\mu$m in diameter and the smallest one is about 34 $\mu$m in diameter. These defects may cause the laser transverse shaping to fail to some extent. And the causes of these defects need further investigation.

These defects are also observed in the SEM images. The SEM image of the biggest defects is shown in Fig. 5(b), which is much smaller than that observed on the QE image. The {\it in situ} imaging can provide additional information obviously.

Additionally, some field emission defects were observed on the surface of cathode \#4, as shown in Fig. 5(c). The QE distribution of cathode \#4 exhibits a step-like shape, which might be caused by some preparation process errors. In addition, there was an obvious field emission defect below the center, which will lead to an increase in dark current. Marking the location and the strength of these field emission defects on the cathode surface will make it possible to find ways to clean them.

The linear optical system still works when the cathode is unloaded, and high-voltage breakdown could be observed when the gun was undergoing a high-voltage conditioning, as shown in Fig. 5(d). The spring structures of the cathode holder in the gun can be observed and it appears to be a good way to monitor the conditioning process.

%because these structures are not smooth enough.

Furthermore, the linear optical system can be designed inside the photocathode preparation chambers with a single solenoid and a YAG screen. In such a case, the changes of QE on the cathode surface can be captured in real-time during its fabrication and preparation.

In summary, an {\it in situ} high resolution real-time  QE imaging system for photocathode was developed in the CAEP THz FEL beamline. The resolution of this system reached about 17 $\mu$m/pixel and the dynamic range was about 24 dB. The dynamic processes of the QE degradation induced by the back-bombardment and temperature rising were observed separately. The defects of the lower QE and field emission were located and measured. Additionally, the breakdown of the gun conditioning could also be captured with this system. This work greatly extends the observing capability in measuring the photocathode performance during preparation and operation, and greatly expands the understanding of the photocathode QE degradation process and details of the preparation process, which in turn benefits researches in high-brightness electron sources, image intensifiers, photo-multipliers tubes and particles accelerators in general.

\bibliography{QEimage.bib}

\begin{thebibliography}{36}%
\makeatletter
\providecommand \@ifxundefined [1]{%
 \@ifx{#1\undefined}
}%
\providecommand \@ifnum [1]{%
 \ifnum #1\expandafter \@firstoftwo
 \else \expandafter \@secondoftwo
 \fi
}%
\providecommand \@ifx [1]{%
 \ifx #1\expandafter \@firstoftwo
 \else \expandafter \@secondoftwo
 \fi
}%
\providecommand \natexlab [1]{#1}%
\providecommand \enquote  [1]{``#1''}%
\providecommand \bibnamefont  [1]{#1}%
\providecommand \bibfnamefont [1]{#1}%
\providecommand \citenamefont [1]{#1}%
\providecommand \href@noop [0]{\@secondoftwo}%
\providecommand \href [0]{\begingroup \@sanitize@url \@href}%
\providecommand \@href[1]{\@@startlink{#1}\@@href}%
\providecommand \@@href[1]{\endgroup#1\@@endlink}%
\providecommand \@sanitize@url [0]{\catcode `\\12\catcode `\$12\catcode
  `\&12\catcode `\#12\catcode `\^12\catcode `\_12\catcode `\%12\relax}%
\providecommand \@@startlink[1]{}%
\providecommand \@@endlink[0]{}%
\providecommand \url  [0]{\begingroup\@sanitize@url \@url }%
\providecommand \@url [1]{\endgroup\@href {#1}{\urlprefix }}%
\providecommand \urlprefix  [0]{URL }%
\providecommand \Eprint [0]{\href }%
\providecommand \doibase [0]{http://dx.doi.org/}%
\providecommand \selectlanguage [0]{\@gobble}%
\providecommand \bibinfo  [0]{\@secondoftwo}%
\providecommand \bibfield  [0]{\@secondoftwo}%
\providecommand \translation [1]{[#1]}%
\providecommand \BibitemOpen [0]{}%
\providecommand \bibitemStop [0]{}%
\providecommand \bibitemNoStop [0]{.\EOS\space}%
\providecommand \EOS [0]{\spacefactor3000\relax}%
\providecommand \BibitemShut  [1]{\csname bibitem#1\endcsname}%
\let\auto@bib@innerbib\@empty
%</preamble>
\bibitem [{\citenamefont {Biberman}(2013)}]{biberman2013photoelectronic}%
  \BibitemOpen
  \bibfield  {author} {\bibinfo {author} {\bibfnamefont {L.}~\bibnamefont
  {Biberman}},\ }\href@noop {} {\emph {\bibinfo {title} {Photoelectronic
  Imaging Devices: Physical processes and methods of analysis}}},\
  Vol.~\bibinfo {volume} {1}\ (\bibinfo  {publisher} {Springer Science \&
  Business Media},\ \bibinfo {year} {2013})\BibitemShut {NoStop}%
\bibitem [{\citenamefont {Kassier}\ \emph {et~al.}(2010)\citenamefont
  {Kassier}, \citenamefont {Haupt}, \citenamefont {Erasmus}, \citenamefont
  {Rohwer}, \citenamefont {Von~Bergmann}, \citenamefont {Schwoerer},
  \citenamefont {Coelho},\ and\ \citenamefont {Auret}}]{kassier2010compact}%
  \BibitemOpen
  \bibfield  {author} {\bibinfo {author} {\bibfnamefont {G.~H.}\ \bibnamefont
  {Kassier}}, \bibinfo {author} {\bibfnamefont {K.}~\bibnamefont {Haupt}},
  \bibinfo {author} {\bibfnamefont {N.}~\bibnamefont {Erasmus}}, \bibinfo
  {author} {\bibfnamefont {E.}~\bibnamefont {Rohwer}}, \bibinfo {author}
  {\bibfnamefont {H.}~\bibnamefont {Von~Bergmann}}, \bibinfo {author}
  {\bibfnamefont {H.}~\bibnamefont {Schwoerer}}, \bibinfo {author}
  {\bibfnamefont {S.~M.}\ \bibnamefont {Coelho}}, \ and\ \bibinfo {author}
  {\bibfnamefont {F.~D.}\ \bibnamefont {Auret}},\ }\href@noop {} {\bibfield
  {journal} {\bibinfo  {journal} {Review of Scientific Instruments}\ }\textbf
  {\bibinfo {volume} {81}},\ \bibinfo {pages} {105103} (\bibinfo {year}
  {2010})}\BibitemShut {NoStop}%
\bibitem [{\citenamefont {Engstrom}(1980)}]{engstrom1980photomultiplier}%
  \BibitemOpen
  \bibfield  {author} {\bibinfo {author} {\bibfnamefont {R.~W.}\ \bibnamefont
  {Engstrom}},\ }\href@noop {} {\emph {\bibinfo {title} {Photomultiplier
  handbook}}}\ (\bibinfo  {publisher} {RCA Corp.},\ \bibinfo {year}
  {1980})\BibitemShut {NoStop}%
\bibitem [{\citenamefont {Dowell}\ \emph {et~al.}(1993)\citenamefont {Dowell},
  \citenamefont {Davis}, \citenamefont {Friddell}, \citenamefont {Tyson},
  \citenamefont {Lancaster}, \citenamefont {Milliman}, \citenamefont
  {Rodenburg}, \citenamefont {Aas}, \citenamefont {Bemes}, \citenamefont
  {Bethel} \emph {et~al.}}]{dowell1993first}%
  \BibitemOpen
  \bibfield  {author} {\bibinfo {author} {\bibfnamefont {D.}~\bibnamefont
  {Dowell}}, \bibinfo {author} {\bibfnamefont {K.}~\bibnamefont {Davis}},
  \bibinfo {author} {\bibfnamefont {K.}~\bibnamefont {Friddell}}, \bibinfo
  {author} {\bibfnamefont {E.}~\bibnamefont {Tyson}}, \bibinfo {author}
  {\bibfnamefont {C.}~\bibnamefont {Lancaster}}, \bibinfo {author}
  {\bibfnamefont {L.}~\bibnamefont {Milliman}}, \bibinfo {author}
  {\bibfnamefont {R.}~\bibnamefont {Rodenburg}}, \bibinfo {author}
  {\bibfnamefont {T.}~\bibnamefont {Aas}}, \bibinfo {author} {\bibfnamefont
  {M.}~\bibnamefont {Bemes}}, \bibinfo {author} {\bibfnamefont
  {S.}~\bibnamefont {Bethel}},  \emph {et~al.},\ }\href@noop {} {\bibfield
  {journal} {\bibinfo  {journal} {Applied physics letters}\ }\textbf {\bibinfo
  {volume} {63}},\ \bibinfo {pages} {2035} (\bibinfo {year}
  {1993})}\BibitemShut {NoStop}%
\bibitem [{\citenamefont {Emma}\ \emph {et~al.}(2010)\citenamefont {Emma},
  \citenamefont {Akre}, \citenamefont {Arthur}, \citenamefont {Bionta},
  \citenamefont {Bostedt}, \citenamefont {Bozek}, \citenamefont {Brachmann},
  \citenamefont {Bucksbaum}, \citenamefont {Coffee}, \citenamefont {Decker}
  \emph {et~al.}}]{emma2010first}%
  \BibitemOpen
  \bibfield  {author} {\bibinfo {author} {\bibfnamefont {P.}~\bibnamefont
  {Emma}}, \bibinfo {author} {\bibfnamefont {R.}~\bibnamefont {Akre}}, \bibinfo
  {author} {\bibfnamefont {J.}~\bibnamefont {Arthur}}, \bibinfo {author}
  {\bibfnamefont {R.}~\bibnamefont {Bionta}}, \bibinfo {author} {\bibfnamefont
  {C.}~\bibnamefont {Bostedt}}, \bibinfo {author} {\bibfnamefont
  {J.}~\bibnamefont {Bozek}}, \bibinfo {author} {\bibfnamefont
  {A.}~\bibnamefont {Brachmann}}, \bibinfo {author} {\bibfnamefont
  {P.}~\bibnamefont {Bucksbaum}}, \bibinfo {author} {\bibfnamefont
  {R.}~\bibnamefont {Coffee}}, \bibinfo {author} {\bibfnamefont {F.-J.}\
  \bibnamefont {Decker}},  \emph {et~al.},\ }\href@noop {} {\bibfield
  {journal} {\bibinfo  {journal} {Nature Photonics}\ }\textbf {\bibinfo
  {volume} {4}},\ \bibinfo {pages} {641} (\bibinfo {year} {2010})}\BibitemShut
  {NoStop}%
\bibitem [{\citenamefont {Dowell}\ \emph {et~al.}(2010)\citenamefont {Dowell},
  \citenamefont {Bazarov}, \citenamefont {Dunham}, \citenamefont {Harkay},
  \citenamefont {Hernandez-Garcia}, \citenamefont {Legg}, \citenamefont
  {Padmore}, \citenamefont {Rao}, \citenamefont {Smedley},\ and\ \citenamefont
  {Wan}}]{dowell2010cathode}%
  \BibitemOpen
  \bibfield  {author} {\bibinfo {author} {\bibfnamefont {D.}~\bibnamefont
  {Dowell}}, \bibinfo {author} {\bibfnamefont {I.}~\bibnamefont {Bazarov}},
  \bibinfo {author} {\bibfnamefont {B.}~\bibnamefont {Dunham}}, \bibinfo
  {author} {\bibfnamefont {K.}~\bibnamefont {Harkay}}, \bibinfo {author}
  {\bibfnamefont {C.}~\bibnamefont {Hernandez-Garcia}}, \bibinfo {author}
  {\bibfnamefont {R.}~\bibnamefont {Legg}}, \bibinfo {author} {\bibfnamefont
  {H.}~\bibnamefont {Padmore}}, \bibinfo {author} {\bibfnamefont
  {T.}~\bibnamefont {Rao}}, \bibinfo {author} {\bibfnamefont {J.}~\bibnamefont
  {Smedley}}, \ and\ \bibinfo {author} {\bibfnamefont {W.}~\bibnamefont
  {Wan}},\ }\href@noop {} {\bibfield  {journal} {\bibinfo  {journal} {Nuclear
  Instruments and Methods in Physics Research Section A: Accelerators,
  Spectrometers, Detectors and Associated Equipment}\ }\textbf {\bibinfo
  {volume} {622}},\ \bibinfo {pages} {685} (\bibinfo {year}
  {2010})}\BibitemShut {NoStop}%
\bibitem [{\citenamefont {Dowell}\ and\ \citenamefont
  {Schmerge}(2009)}]{Dowell2009}%
  \BibitemOpen
  \bibfield  {author} {\bibinfo {author} {\bibfnamefont {D.~H.}\ \bibnamefont
  {Dowell}}\ and\ \bibinfo {author} {\bibfnamefont {J.~F.}\ \bibnamefont
  {Schmerge}},\ }\href {\doibase 10.1103/PhysRevSTAB.12.074201} {\bibfield
  {journal} {\bibinfo  {journal} {Physical Review Special Topics - Accelerators
  and Beams}\ }\textbf {\bibinfo {volume} {12}},\ \bibinfo {pages} {074201}
  (\bibinfo {year} {2009})}\BibitemShut {NoStop}%
\bibitem [{\citenamefont {Akre}\ \emph {et~al.}(2008)\citenamefont {Akre},
  \citenamefont {Dowell}, \citenamefont {Emma}, \citenamefont {Frisch},
  \citenamefont {Gilevich}, \citenamefont {Hays}, \citenamefont {Hering},
  \citenamefont {Iverson}, \citenamefont {Limborg-Deprey}, \citenamefont
  {Loos}, \citenamefont {Miahnahri}, \citenamefont {Schmerge}, \citenamefont
  {Turner}, \citenamefont {Welch}, \citenamefont {White},\ and\ \citenamefont
  {Wu}}]{PhysRevSTAB.11.030703}%
  \BibitemOpen
  \bibfield  {author} {\bibinfo {author} {\bibfnamefont {R.}~\bibnamefont
  {Akre}}, \bibinfo {author} {\bibfnamefont {D.}~\bibnamefont {Dowell}},
  \bibinfo {author} {\bibfnamefont {P.}~\bibnamefont {Emma}}, \bibinfo {author}
  {\bibfnamefont {J.}~\bibnamefont {Frisch}}, \bibinfo {author} {\bibfnamefont
  {S.}~\bibnamefont {Gilevich}}, \bibinfo {author} {\bibfnamefont
  {G.}~\bibnamefont {Hays}}, \bibinfo {author} {\bibfnamefont {P.}~\bibnamefont
  {Hering}}, \bibinfo {author} {\bibfnamefont {R.}~\bibnamefont {Iverson}},
  \bibinfo {author} {\bibfnamefont {C.}~\bibnamefont {Limborg-Deprey}},
  \bibinfo {author} {\bibfnamefont {H.}~\bibnamefont {Loos}}, \bibinfo {author}
  {\bibfnamefont {A.}~\bibnamefont {Miahnahri}}, \bibinfo {author}
  {\bibfnamefont {J.}~\bibnamefont {Schmerge}}, \bibinfo {author}
  {\bibfnamefont {J.}~\bibnamefont {Turner}}, \bibinfo {author} {\bibfnamefont
  {J.}~\bibnamefont {Welch}}, \bibinfo {author} {\bibfnamefont
  {W.}~\bibnamefont {White}}, \ and\ \bibinfo {author} {\bibfnamefont
  {J.}~\bibnamefont {Wu}},\ }\href {\doibase 10.1103/PhysRevSTAB.11.030703}
  {\bibfield  {journal} {\bibinfo  {journal} {Phys. Rev. ST Accel. Beams}\
  }\textbf {\bibinfo {volume} {11}},\ \bibinfo {pages} {030703} (\bibinfo
  {year} {2008})}\BibitemShut {NoStop}%
\bibitem [{\citenamefont {Ding}\ \emph {et~al.}(2009)\citenamefont {Ding},
  \citenamefont {Brachmann}, \citenamefont {Decker}, \citenamefont {Dowell},
  \citenamefont {Emma}, \citenamefont {Frisch}, \citenamefont {Gilevich},
  \citenamefont {Hays}, \citenamefont {Hering}, \citenamefont {Huang} \emph
  {et~al.}}]{ding2009measurements}%
  \BibitemOpen
  \bibfield  {author} {\bibinfo {author} {\bibfnamefont {Y.}~\bibnamefont
  {Ding}}, \bibinfo {author} {\bibfnamefont {A.}~\bibnamefont {Brachmann}},
  \bibinfo {author} {\bibfnamefont {F.-J.}\ \bibnamefont {Decker}}, \bibinfo
  {author} {\bibfnamefont {D.}~\bibnamefont {Dowell}}, \bibinfo {author}
  {\bibfnamefont {P.}~\bibnamefont {Emma}}, \bibinfo {author} {\bibfnamefont
  {J.}~\bibnamefont {Frisch}}, \bibinfo {author} {\bibfnamefont
  {S.}~\bibnamefont {Gilevich}}, \bibinfo {author} {\bibfnamefont
  {G.}~\bibnamefont {Hays}}, \bibinfo {author} {\bibfnamefont {P.}~\bibnamefont
  {Hering}}, \bibinfo {author} {\bibfnamefont {Z.}~\bibnamefont {Huang}},
  \emph {et~al.},\ }\href@noop {} {\bibfield  {journal} {\bibinfo  {journal}
  {Physical review letters}\ }\textbf {\bibinfo {volume} {102}},\ \bibinfo
  {pages} {254801} (\bibinfo {year} {2009})}\BibitemShut {NoStop}%
\bibitem [{\citenamefont {Zheng}\ \emph {et~al.}(2016)\citenamefont {Zheng},
  \citenamefont {Du}, \citenamefont {Zhang}, \citenamefont {Qian},
  \citenamefont {Yan}, \citenamefont {Shi}, \citenamefont {Zhang},
  \citenamefont {Zhou}, \citenamefont {Wu}, \citenamefont {Su}, \citenamefont
  {Wang}, \citenamefont {Tian}, \citenamefont {Huang}, \citenamefont {Chen},\
  and\ \citenamefont {Tang}}]{Zheng201698}%
  \BibitemOpen
  \bibfield  {author} {\bibinfo {author} {\bibfnamefont {L.}~\bibnamefont
  {Zheng}}, \bibinfo {author} {\bibfnamefont {Y.}~\bibnamefont {Du}}, \bibinfo
  {author} {\bibfnamefont {Z.}~\bibnamefont {Zhang}}, \bibinfo {author}
  {\bibfnamefont {H.}~\bibnamefont {Qian}}, \bibinfo {author} {\bibfnamefont
  {L.}~\bibnamefont {Yan}}, \bibinfo {author} {\bibfnamefont {J.}~\bibnamefont
  {Shi}}, \bibinfo {author} {\bibfnamefont {Z.}~\bibnamefont {Zhang}}, \bibinfo
  {author} {\bibfnamefont {Z.}~\bibnamefont {Zhou}}, \bibinfo {author}
  {\bibfnamefont {X.}~\bibnamefont {Wu}}, \bibinfo {author} {\bibfnamefont
  {X.}~\bibnamefont {Su}}, \bibinfo {author} {\bibfnamefont {D.}~\bibnamefont
  {Wang}}, \bibinfo {author} {\bibfnamefont {Q.}~\bibnamefont {Tian}}, \bibinfo
  {author} {\bibfnamefont {W.}~\bibnamefont {Huang}}, \bibinfo {author}
  {\bibfnamefont {H.}~\bibnamefont {Chen}}, \ and\ \bibinfo {author}
  {\bibfnamefont {C.}~\bibnamefont {Tang}},\ }\href {\doibase
  http://dx.doi.org/10.1016/j.nima.2016.07.015} {\bibfield  {journal} {\bibinfo
   {journal} {Nuclear Instruments and Methods in Physics Research Section A:
  Accelerators, Spectrometers, Detectors and Associated Equipment}\ }\textbf
  {\bibinfo {volume} {834}},\ \bibinfo {pages} {98 } (\bibinfo {year}
  {2016})}\BibitemShut {NoStop}%
\bibitem [{\citenamefont {Chapman}(2009)}]{chapman2009x}%
  \BibitemOpen
  \bibfield  {author} {\bibinfo {author} {\bibfnamefont {H.~N.}\ \bibnamefont
  {Chapman}},\ }\href@noop {} {\bibfield  {journal} {\bibinfo  {journal}
  {Nature materials}\ }\textbf {\bibinfo {volume} {8}},\ \bibinfo {pages} {299}
  (\bibinfo {year} {2009})}\BibitemShut {NoStop}%
\bibitem [{\citenamefont {Rohringer}\ \emph {et~al.}(2012)\citenamefont
  {Rohringer}, \citenamefont {Ryan}, \citenamefont {London}, \citenamefont
  {Purvis}, \citenamefont {Albert}, \citenamefont {Dunn}, \citenamefont
  {Bozek}, \citenamefont {Bostedt}, \citenamefont {Graf}, \citenamefont {Hill}
  \emph {et~al.}}]{rohringer2012atomic}%
  \BibitemOpen
  \bibfield  {author} {\bibinfo {author} {\bibfnamefont {N.}~\bibnamefont
  {Rohringer}}, \bibinfo {author} {\bibfnamefont {D.}~\bibnamefont {Ryan}},
  \bibinfo {author} {\bibfnamefont {R.~A.}\ \bibnamefont {London}}, \bibinfo
  {author} {\bibfnamefont {M.}~\bibnamefont {Purvis}}, \bibinfo {author}
  {\bibfnamefont {F.}~\bibnamefont {Albert}}, \bibinfo {author} {\bibfnamefont
  {J.}~\bibnamefont {Dunn}}, \bibinfo {author} {\bibfnamefont {J.~D.}\
  \bibnamefont {Bozek}}, \bibinfo {author} {\bibfnamefont {C.}~\bibnamefont
  {Bostedt}}, \bibinfo {author} {\bibfnamefont {A.}~\bibnamefont {Graf}},
  \bibinfo {author} {\bibfnamefont {R.}~\bibnamefont {Hill}},  \emph {et~al.},\
  }\href@noop {} {\bibfield  {journal} {\bibinfo  {journal} {Nature}\ }\textbf
  {\bibinfo {volume} {481}},\ \bibinfo {pages} {488} (\bibinfo {year}
  {2012})}\BibitemShut {NoStop}%
\bibitem [{\citenamefont {Hirata}\ \emph {et~al.}(2014)\citenamefont {Hirata},
  \citenamefont {Shinzawa-Itoh}, \citenamefont {Yano}, \citenamefont
  {Takemura}, \citenamefont {Kato}, \citenamefont {Hatanaka}, \citenamefont
  {Muramoto}, \citenamefont {Kawahara}, \citenamefont {Tsukihara},
  \citenamefont {Yamashita} \emph {et~al.}}]{hirata2014determination}%
  \BibitemOpen
  \bibfield  {author} {\bibinfo {author} {\bibfnamefont {K.}~\bibnamefont
  {Hirata}}, \bibinfo {author} {\bibfnamefont {K.}~\bibnamefont
  {Shinzawa-Itoh}}, \bibinfo {author} {\bibfnamefont {N.}~\bibnamefont {Yano}},
  \bibinfo {author} {\bibfnamefont {S.}~\bibnamefont {Takemura}}, \bibinfo
  {author} {\bibfnamefont {K.}~\bibnamefont {Kato}}, \bibinfo {author}
  {\bibfnamefont {M.}~\bibnamefont {Hatanaka}}, \bibinfo {author}
  {\bibfnamefont {K.}~\bibnamefont {Muramoto}}, \bibinfo {author}
  {\bibfnamefont {T.}~\bibnamefont {Kawahara}}, \bibinfo {author}
  {\bibfnamefont {T.}~\bibnamefont {Tsukihara}}, \bibinfo {author}
  {\bibfnamefont {E.}~\bibnamefont {Yamashita}},  \emph {et~al.},\ }\href@noop
  {} {\bibfield  {journal} {\bibinfo  {journal} {Nature methods}\ }\textbf
  {\bibinfo {volume} {11}},\ \bibinfo {pages} {734} (\bibinfo {year}
  {2014})}\BibitemShut {NoStop}%
\bibitem [{\citenamefont {Suga}\ \emph {et~al.}(2015)\citenamefont {Suga},
  \citenamefont {Akita}, \citenamefont {Hirata}, \citenamefont {Ueno},
  \citenamefont {Murakami}, \citenamefont {Nakajima}, \citenamefont {Shimizu},
  \citenamefont {Yamashita}, \citenamefont {Yamamoto}, \citenamefont {Ago}
  \emph {et~al.}}]{suga2015native}%
  \BibitemOpen
  \bibfield  {author} {\bibinfo {author} {\bibfnamefont {M.}~\bibnamefont
  {Suga}}, \bibinfo {author} {\bibfnamefont {F.}~\bibnamefont {Akita}},
  \bibinfo {author} {\bibfnamefont {K.}~\bibnamefont {Hirata}}, \bibinfo
  {author} {\bibfnamefont {G.}~\bibnamefont {Ueno}}, \bibinfo {author}
  {\bibfnamefont {H.}~\bibnamefont {Murakami}}, \bibinfo {author}
  {\bibfnamefont {Y.}~\bibnamefont {Nakajima}}, \bibinfo {author}
  {\bibfnamefont {T.}~\bibnamefont {Shimizu}}, \bibinfo {author} {\bibfnamefont
  {K.}~\bibnamefont {Yamashita}}, \bibinfo {author} {\bibfnamefont
  {M.}~\bibnamefont {Yamamoto}}, \bibinfo {author} {\bibfnamefont
  {H.}~\bibnamefont {Ago}},  \emph {et~al.},\ }\href@noop {} {\bibfield
  {journal} {\bibinfo  {journal} {Nature}\ }\textbf {\bibinfo {volume} {517}},\
  \bibinfo {pages} {99} (\bibinfo {year} {2015})}\BibitemShut {NoStop}%
\bibitem [{\citenamefont {Fischer}\ \emph {et~al.}(1994)\citenamefont
  {Fischer}, \citenamefont {Drescher},\ and\ \citenamefont
  {Reichert}}]{fischer1994thermal}%
  \BibitemOpen
  \bibfield  {author} {\bibinfo {author} {\bibfnamefont {H.}~\bibnamefont
  {Fischer}}, \bibinfo {author} {\bibfnamefont {P.}~\bibnamefont {Drescher}}, \
  and\ \bibinfo {author} {\bibfnamefont {E.}~\bibnamefont {Reichert}},\ }in\
  \href@noop {} {\emph {\bibinfo {booktitle} {Photocathodes for Polarized
  Electron Sources for Accelerators}}}\ (\bibinfo {year} {1994})\ pp.\ \bibinfo
  {pages} {249--260}\BibitemShut {NoStop}%
\bibitem [{\citenamefont {Aulenbacher}\ \emph {et~al.}(1997)\citenamefont
  {Aulenbacher}, \citenamefont {Nachtigall}, \citenamefont {Andresen},
  \citenamefont {Bermuth}, \citenamefont {Dombo}, \citenamefont {Drescher},
  \citenamefont {Euteneuer}, \citenamefont {Fischer}, \citenamefont {Harrach},
  \citenamefont {Hartmann}, \citenamefont {Hoffmann}, \citenamefont
  {Jennewein}, \citenamefont {Kaiser}, \citenamefont {K枚bis}, \citenamefont
  {Kreidel}, \citenamefont {Langbein}, \citenamefont {Petri}, \citenamefont
  {Pl眉tzer}, \citenamefont {Reichert}, \citenamefont {Schemies},
  \citenamefont {Sch枚pe}, \citenamefont {Steffens}, \citenamefont
  {Steigerwald}, \citenamefont {Trautner},\ and\ \citenamefont
  {Weis}}]{Aulenbacher1997498}%
  \BibitemOpen
  \bibfield  {author} {\bibinfo {author} {\bibfnamefont {K.}~\bibnamefont
  {Aulenbacher}}, \bibinfo {author} {\bibfnamefont {C.}~\bibnamefont
  {Nachtigall}}, \bibinfo {author} {\bibfnamefont {H.}~\bibnamefont
  {Andresen}}, \bibinfo {author} {\bibfnamefont {J.}~\bibnamefont {Bermuth}},
  \bibinfo {author} {\bibfnamefont {T.}~\bibnamefont {Dombo}}, \bibinfo
  {author} {\bibfnamefont {P.}~\bibnamefont {Drescher}}, \bibinfo {author}
  {\bibfnamefont {H.}~\bibnamefont {Euteneuer}}, \bibinfo {author}
  {\bibfnamefont {H.}~\bibnamefont {Fischer}}, \bibinfo {author} {\bibfnamefont
  {D.}~\bibnamefont {Harrach}}, \bibinfo {author} {\bibfnamefont
  {P.}~\bibnamefont {Hartmann}}, \bibinfo {author} {\bibfnamefont
  {J.}~\bibnamefont {Hoffmann}}, \bibinfo {author} {\bibfnamefont
  {P.}~\bibnamefont {Jennewein}}, \bibinfo {author} {\bibfnamefont
  {K.}~\bibnamefont {Kaiser}}, \bibinfo {author} {\bibfnamefont
  {S.}~\bibnamefont {K枚bis}}, \bibinfo {author} {\bibfnamefont
  {H.}~\bibnamefont {Kreidel}}, \bibinfo {author} {\bibfnamefont
  {J.}~\bibnamefont {Langbein}}, \bibinfo {author} {\bibfnamefont
  {M.}~\bibnamefont {Petri}}, \bibinfo {author} {\bibfnamefont
  {S.}~\bibnamefont {Pl眉tzer}}, \bibinfo {author} {\bibfnamefont
  {E.}~\bibnamefont {Reichert}}, \bibinfo {author} {\bibfnamefont
  {M.}~\bibnamefont {Schemies}}, \bibinfo {author} {\bibfnamefont {H.-J.}\
  \bibnamefont {Sch枚pe}}, \bibinfo {author} {\bibfnamefont {K.-H.}\
  \bibnamefont {Steffens}}, \bibinfo {author} {\bibfnamefont {M.}~\bibnamefont
  {Steigerwald}}, \bibinfo {author} {\bibfnamefont {H.}~\bibnamefont
  {Trautner}}, \ and\ \bibinfo {author} {\bibfnamefont {T.}~\bibnamefont
  {Weis}},\ }\href {\doibase http://dx.doi.org/10.1016/S0168-9002(97)00528-7}
  {\bibfield  {journal} {\bibinfo  {journal} {Nuclear Instruments and Methods
  in Physics Research Section A: Accelerators, Spectrometers, Detectors and
  Associated Equipment}\ }\textbf {\bibinfo {volume} {391}},\ \bibinfo {pages}
  {498 } (\bibinfo {year} {1997})}\BibitemShut {NoStop}%
\bibitem [{\citenamefont {Lewellen}(2002)}]{lewellen2002ion}%
  \BibitemOpen
  \bibfield  {author} {\bibinfo {author} {\bibfnamefont {J.~W.}\ \bibnamefont
  {Lewellen}},\ }\href@noop {} {\bibfield  {journal} {\bibinfo  {journal}
  {Physical Review Special Topics-Accelerators and Beams}\ }\textbf {\bibinfo
  {volume} {5}},\ \bibinfo {pages} {020101} (\bibinfo {year}
  {2002})}\BibitemShut {NoStop}%
\bibitem [{\citenamefont {Qiang}(2010)}]{qiang2010particle}%
  \BibitemOpen
  \bibfield  {author} {\bibinfo {author} {\bibfnamefont {J.}~\bibnamefont
  {Qiang}},\ }\href@noop {} {\bibfield  {journal} {\bibinfo  {journal} {Nuclear
  Instruments and Methods in Physics Research Section A: Accelerators,
  Spectrometers, Detectors and Associated Equipment}\ }\textbf {\bibinfo
  {volume} {614}},\ \bibinfo {pages} {1} (\bibinfo {year} {2010})}\BibitemShut
  {NoStop}%
\bibitem [{\citenamefont {Zhang}\ \emph {et~al.}(2011)\citenamefont {Zhang},
  \citenamefont {Benson},\ and\ \citenamefont {H-Garcia}}]{Zhang2011a}%
  \BibitemOpen
  \bibfield  {author} {\bibinfo {author} {\bibfnamefont {S.}~\bibnamefont
  {Zhang}}, \bibinfo {author} {\bibfnamefont {S.}~\bibnamefont {Benson}}, \
  and\ \bibinfo {author} {\bibfnamefont {C.}~\bibnamefont {H-Garcia}},\ }\href
  {\doibase http://dx.doi.org/10.1016/j.nima.2010.12.132} {\bibfield  {journal}
  {\bibinfo  {journal} {Nuclear Instruments and Methods in Physics Research
  Section A: Accelerators, Spectrometers, Detectors and Associated Equipment}\
  }\textbf {\bibinfo {volume} {631}},\ \bibinfo {pages} {22 } (\bibinfo {year}
  {2011})}\BibitemShut {NoStop}%
\bibitem [{\citenamefont {Grames}\ \emph {et~al.}(2011)\citenamefont {Grames},
  \citenamefont {Suleiman}, \citenamefont {Adderley}, \citenamefont {Clark},
  \citenamefont {Hansknecht}, \citenamefont {Machie}, \citenamefont {Poelker},\
  and\ \citenamefont {Stutzman}}]{grames2011charge}%
  \BibitemOpen
  \bibfield  {author} {\bibinfo {author} {\bibfnamefont {J.}~\bibnamefont
  {Grames}}, \bibinfo {author} {\bibfnamefont {R.}~\bibnamefont {Suleiman}},
  \bibinfo {author} {\bibfnamefont {P.}~\bibnamefont {Adderley}}, \bibinfo
  {author} {\bibfnamefont {J.}~\bibnamefont {Clark}}, \bibinfo {author}
  {\bibfnamefont {J.}~\bibnamefont {Hansknecht}}, \bibinfo {author}
  {\bibfnamefont {D.}~\bibnamefont {Machie}}, \bibinfo {author} {\bibfnamefont
  {M.}~\bibnamefont {Poelker}}, \ and\ \bibinfo {author} {\bibfnamefont
  {M.}~\bibnamefont {Stutzman}},\ }\href@noop {} {\bibfield  {journal}
  {\bibinfo  {journal} {Physical Review Special Topics-Accelerators and Beams}\
  }\textbf {\bibinfo {volume} {14}},\ \bibinfo {pages} {043501} (\bibinfo
  {year} {2011})}\BibitemShut {NoStop}%
\bibitem [{\citenamefont {Pandey}\ \emph {et~al.}(2009)\citenamefont {Pandey},
  \citenamefont {M{\"u}ller}, \citenamefont {Reschke},\ and\ \citenamefont
  {Singer}}]{pandey2009field}%
  \BibitemOpen
  \bibfield  {author} {\bibinfo {author} {\bibfnamefont {A.~D.}\ \bibnamefont
  {Pandey}}, \bibinfo {author} {\bibfnamefont {G.}~\bibnamefont {M{\"u}ller}},
  \bibinfo {author} {\bibfnamefont {D.}~\bibnamefont {Reschke}}, \ and\
  \bibinfo {author} {\bibfnamefont {X.}~\bibnamefont {Singer}},\ }\href@noop {}
  {\bibfield  {journal} {\bibinfo  {journal} {Physical Review Special
  Topics-Accelerators and Beams}\ }\textbf {\bibinfo {volume} {12}},\ \bibinfo
  {pages} {023501} (\bibinfo {year} {2009})}\BibitemShut {NoStop}%
\bibitem [{\citenamefont {Shao}\ \emph {et~al.}(2016)\citenamefont {Shao},
  \citenamefont {Shi}, \citenamefont {Antipov}, \citenamefont {Baryshev},
  \citenamefont {Chen}, \citenamefont {Conde}, \citenamefont {Gai},
  \citenamefont {Ha}, \citenamefont {Jing}, \citenamefont {Wang} \emph
  {et~al.}}]{shao2016situ}%
  \BibitemOpen
  \bibfield  {author} {\bibinfo {author} {\bibfnamefont {J.}~\bibnamefont
  {Shao}}, \bibinfo {author} {\bibfnamefont {J.}~\bibnamefont {Shi}}, \bibinfo
  {author} {\bibfnamefont {S.~P.}\ \bibnamefont {Antipov}}, \bibinfo {author}
  {\bibfnamefont {S.~V.}\ \bibnamefont {Baryshev}}, \bibinfo {author}
  {\bibfnamefont {H.}~\bibnamefont {Chen}}, \bibinfo {author} {\bibfnamefont
  {M.}~\bibnamefont {Conde}}, \bibinfo {author} {\bibfnamefont
  {W.}~\bibnamefont {Gai}}, \bibinfo {author} {\bibfnamefont {G.}~\bibnamefont
  {Ha}}, \bibinfo {author} {\bibfnamefont {C.}~\bibnamefont {Jing}}, \bibinfo
  {author} {\bibfnamefont {F.}~\bibnamefont {Wang}},  \emph {et~al.},\
  }\href@noop {} {\bibfield  {journal} {\bibinfo  {journal} {Physical review
  letters}\ }\textbf {\bibinfo {volume} {117}},\ \bibinfo {pages} {084801}
  (\bibinfo {year} {2016})}\BibitemShut {NoStop}%
\bibitem [{\citenamefont {Gubeli}\ \emph {et~al.}(2001)\citenamefont {Gubeli},
  \citenamefont {Evans}, \citenamefont {Grippo}, \citenamefont {Jordan},
  \citenamefont {Shinn},\ and\ \citenamefont {Siggins}}]{Gubeli2001554}%
  \BibitemOpen
  \bibfield  {author} {\bibinfo {author} {\bibfnamefont {J.}~\bibnamefont
  {Gubeli}}, \bibinfo {author} {\bibfnamefont {R.}~\bibnamefont {Evans}},
  \bibinfo {author} {\bibfnamefont {A.}~\bibnamefont {Grippo}}, \bibinfo
  {author} {\bibfnamefont {K.}~\bibnamefont {Jordan}}, \bibinfo {author}
  {\bibfnamefont {M.}~\bibnamefont {Shinn}}, \ and\ \bibinfo {author}
  {\bibfnamefont {T.}~\bibnamefont {Siggins}},\ }\href {\doibase
  http://dx.doi.org/10.1016/S0168-9002(01)01695-3} {\bibfield  {journal}
  {\bibinfo  {journal} {Nuclear Instruments and Methods in Physics Research
  Section A: Accelerators, Spectrometers, Detectors and Associated Equipment}\
  }\textbf {\bibinfo {volume} {475}},\ \bibinfo {pages} {554 } (\bibinfo {year}
  {2001})}\BibitemShut {NoStop}%
\bibitem [{\citenamefont {Sinclair}\ \emph {et~al.}(2007)\citenamefont
  {Sinclair}, \citenamefont {Adderley}, \citenamefont {Dunham}, \citenamefont
  {Hansknecht}, \citenamefont {Hartmann}, \citenamefont {Poelker},
  \citenamefont {Price}, \citenamefont {Rutt}, \citenamefont {Schneider},\ and\
  \citenamefont {Steigerwald}}]{sinclair2007development}%
  \BibitemOpen
  \bibfield  {author} {\bibinfo {author} {\bibfnamefont {C.}~\bibnamefont
  {Sinclair}}, \bibinfo {author} {\bibfnamefont {P.}~\bibnamefont {Adderley}},
  \bibinfo {author} {\bibfnamefont {B.}~\bibnamefont {Dunham}}, \bibinfo
  {author} {\bibfnamefont {J.}~\bibnamefont {Hansknecht}}, \bibinfo {author}
  {\bibfnamefont {P.}~\bibnamefont {Hartmann}}, \bibinfo {author}
  {\bibfnamefont {M.}~\bibnamefont {Poelker}}, \bibinfo {author} {\bibfnamefont
  {J.}~\bibnamefont {Price}}, \bibinfo {author} {\bibfnamefont
  {P.}~\bibnamefont {Rutt}}, \bibinfo {author} {\bibfnamefont {W.}~\bibnamefont
  {Schneider}}, \ and\ \bibinfo {author} {\bibfnamefont {M.}~\bibnamefont
  {Steigerwald}},\ }\href@noop {} {\bibfield  {journal} {\bibinfo  {journal}
  {Physical Review Special Topics-Accelerators and Beams}\ }\textbf {\bibinfo
  {volume} {10}},\ \bibinfo {pages} {023501} (\bibinfo {year}
  {2007})}\BibitemShut {NoStop}%
\bibitem [{\citenamefont {Cultrera}\ \emph {et~al.}(2011)\citenamefont
  {Cultrera}, \citenamefont {Bazarov}, \citenamefont {Bartnik}, \citenamefont
  {Dunham}, \citenamefont {Karkare}, \citenamefont {Merluzzi},\ and\
  \citenamefont {Nichols}}]{cultrera2011thermal}%
  \BibitemOpen
  \bibfield  {author} {\bibinfo {author} {\bibfnamefont {L.}~\bibnamefont
  {Cultrera}}, \bibinfo {author} {\bibfnamefont {I.}~\bibnamefont {Bazarov}},
  \bibinfo {author} {\bibfnamefont {A.}~\bibnamefont {Bartnik}}, \bibinfo
  {author} {\bibfnamefont {B.}~\bibnamefont {Dunham}}, \bibinfo {author}
  {\bibfnamefont {S.}~\bibnamefont {Karkare}}, \bibinfo {author} {\bibfnamefont
  {R.}~\bibnamefont {Merluzzi}}, \ and\ \bibinfo {author} {\bibfnamefont
  {M.}~\bibnamefont {Nichols}},\ }\href@noop {} {\bibfield  {journal} {\bibinfo
   {journal} {Applied Physics Letters}\ }\textbf {\bibinfo {volume} {99}},\
  \bibinfo {pages} {152110} (\bibinfo {year} {2011})}\BibitemShut {NoStop}%
\bibitem [{\citenamefont {Zhou}\ \emph {et~al.}(2012)\citenamefont {Zhou},
  \citenamefont {Brachmann}, \citenamefont {Decker}, \citenamefont {Emma},
  \citenamefont {Gilevich}, \citenamefont {Iverson}, \citenamefont {Stefan},\
  and\ \citenamefont {Turner}}]{PhysRevSTAB.15.090703}%
  \BibitemOpen
  \bibfield  {author} {\bibinfo {author} {\bibfnamefont {F.}~\bibnamefont
  {Zhou}}, \bibinfo {author} {\bibfnamefont {A.}~\bibnamefont {Brachmann}},
  \bibinfo {author} {\bibfnamefont {F.-J.}\ \bibnamefont {Decker}}, \bibinfo
  {author} {\bibfnamefont {P.}~\bibnamefont {Emma}}, \bibinfo {author}
  {\bibfnamefont {S.}~\bibnamefont {Gilevich}}, \bibinfo {author}
  {\bibfnamefont {R.}~\bibnamefont {Iverson}}, \bibinfo {author} {\bibfnamefont
  {P.}~\bibnamefont {Stefan}}, \ and\ \bibinfo {author} {\bibfnamefont
  {J.}~\bibnamefont {Turner}},\ }\href {\doibase 10.1103/PhysRevSTAB.15.090703}
  {\bibfield  {journal} {\bibinfo  {journal} {Phys. Rev. ST Accel. Beams}\
  }\textbf {\bibinfo {volume} {15}},\ \bibinfo {pages} {090703} (\bibinfo
  {year} {2012})}\BibitemShut {NoStop}%
\bibitem [{\citenamefont {Riddick}\ \emph {et~al.}(2013)\citenamefont
  {Riddick}, \citenamefont {Montgomery}, \citenamefont {Fiorito}, \citenamefont
  {Zhang}, \citenamefont {Shkvarunets}, \citenamefont {Pan},\ and\
  \citenamefont {Khan}}]{riddick2013photocathode}%
  \BibitemOpen
  \bibfield  {author} {\bibinfo {author} {\bibfnamefont {B.}~\bibnamefont
  {Riddick}}, \bibinfo {author} {\bibfnamefont {E.}~\bibnamefont {Montgomery}},
  \bibinfo {author} {\bibfnamefont {R.}~\bibnamefont {Fiorito}}, \bibinfo
  {author} {\bibfnamefont {H.}~\bibnamefont {Zhang}}, \bibinfo {author}
  {\bibfnamefont {A.}~\bibnamefont {Shkvarunets}}, \bibinfo {author}
  {\bibfnamefont {Z.}~\bibnamefont {Pan}}, \ and\ \bibinfo {author}
  {\bibfnamefont {S.}~\bibnamefont {Khan}},\ }\href@noop {} {\bibfield
  {journal} {\bibinfo  {journal} {Physical Review Special Topics-Accelerators
  and Beams}\ }\textbf {\bibinfo {volume} {16}},\ \bibinfo {pages} {062802}
  (\bibinfo {year} {2013})}\BibitemShut {NoStop}%
\bibitem [{\citenamefont {Filippetto}\ \emph {et~al.}(2015)\citenamefont
  {Filippetto}, \citenamefont {Qian},\ and\ \citenamefont
  {Sannibale}}]{filippetto2015cesium}%
  \BibitemOpen
  \bibfield  {author} {\bibinfo {author} {\bibfnamefont {D.}~\bibnamefont
  {Filippetto}}, \bibinfo {author} {\bibfnamefont {H.}~\bibnamefont {Qian}}, \
  and\ \bibinfo {author} {\bibfnamefont {F.}~\bibnamefont {Sannibale}},\
  }\href@noop {} {\bibfield  {journal} {\bibinfo  {journal} {Applied Physics
  Letters}\ }\textbf {\bibinfo {volume} {107}},\ \bibinfo {pages} {042104}
  (\bibinfo {year} {2015})}\BibitemShut {NoStop}%
\bibitem [{\citenamefont {Anders}\ \emph {et~al.}(1999)\citenamefont {Anders},
  \citenamefont {Padmore}, \citenamefont {Duarte}, \citenamefont {Renner},
  \citenamefont {Stammler}, \citenamefont {Scholl}, \citenamefont {Scheinfein},
  \citenamefont {Stöhr}, \citenamefont {Séve},\ and\ \citenamefont
  {Sinkovic}}]{PEEM1999}%
  \BibitemOpen
  \bibfield  {author} {\bibinfo {author} {\bibfnamefont {S.}~\bibnamefont
  {Anders}}, \bibinfo {author} {\bibfnamefont {H.~A.}\ \bibnamefont {Padmore}},
  \bibinfo {author} {\bibfnamefont {R.~M.}\ \bibnamefont {Duarte}}, \bibinfo
  {author} {\bibfnamefont {T.}~\bibnamefont {Renner}}, \bibinfo {author}
  {\bibfnamefont {T.}~\bibnamefont {Stammler}}, \bibinfo {author}
  {\bibfnamefont {A.}~\bibnamefont {Scholl}}, \bibinfo {author} {\bibfnamefont
  {M.~R.}\ \bibnamefont {Scheinfein}}, \bibinfo {author} {\bibfnamefont
  {J.}~\bibnamefont {Stöhr}}, \bibinfo {author} {\bibfnamefont
  {L.}~\bibnamefont {Séve}}, \ and\ \bibinfo {author} {\bibfnamefont
  {B.}~\bibnamefont {Sinkovic}},\ }\href {\doibase
  http://dx.doi.org/10.1063/1.1150023} {\bibfield  {journal} {\bibinfo
  {journal} {Review of Scientific Instruments}\ }\textbf {\bibinfo {volume}
  {70}},\ \bibinfo {pages} {3973} (\bibinfo {year} {1999})}\BibitemShut
  {NoStop}%
\bibitem [{\citenamefont {Polyakov}\ \emph {et~al.}(2013)\citenamefont
  {Polyakov}, \citenamefont {Senft}, \citenamefont {Thompson}, \citenamefont
  {Feng}, \citenamefont {Cabrini}, \citenamefont {Schuck}, \citenamefont
  {Padmore}, \citenamefont {Peppernick},\ and\ \citenamefont
  {Hess}}]{PhysRevLett.110.076802}%
  \BibitemOpen
  \bibfield  {author} {\bibinfo {author} {\bibfnamefont {A.}~\bibnamefont
  {Polyakov}}, \bibinfo {author} {\bibfnamefont {C.}~\bibnamefont {Senft}},
  \bibinfo {author} {\bibfnamefont {K.~F.}\ \bibnamefont {Thompson}}, \bibinfo
  {author} {\bibfnamefont {J.}~\bibnamefont {Feng}}, \bibinfo {author}
  {\bibfnamefont {S.}~\bibnamefont {Cabrini}}, \bibinfo {author} {\bibfnamefont
  {P.~J.}\ \bibnamefont {Schuck}}, \bibinfo {author} {\bibfnamefont {H.~A.}\
  \bibnamefont {Padmore}}, \bibinfo {author} {\bibfnamefont {S.~J.}\
  \bibnamefont {Peppernick}}, \ and\ \bibinfo {author} {\bibfnamefont {W.~P.}\
  \bibnamefont {Hess}},\ }\href {\doibase 10.1103/PhysRevLett.110.076802}
  {\bibfield  {journal} {\bibinfo  {journal} {Phys. Rev. Lett.}\ }\textbf
  {\bibinfo {volume} {110}},\ \bibinfo {pages} {076802} (\bibinfo {year}
  {2013})}\BibitemShut {NoStop}%
\bibitem [{\citenamefont {Wu}\ \emph {et~al.}(2014)\citenamefont {Wu},
  \citenamefont {Pan}, \citenamefont {Li}, \citenamefont {Xiao}, \citenamefont
  {Yang}, \citenamefont {Wang}, \citenamefont {Yang},\ and\ \citenamefont
  {Li}}]{wudaiSAP2014}%
  \BibitemOpen
  \bibfield  {author} {\bibinfo {author} {\bibfnamefont {D.}~\bibnamefont
  {Wu}}, \bibinfo {author} {\bibfnamefont {Q.}~\bibnamefont {Pan}}, \bibinfo
  {author} {\bibfnamefont {K.}~\bibnamefont {Li}}, \bibinfo {author}
  {\bibfnamefont {D.}~\bibnamefont {Xiao}}, \bibinfo {author} {\bibfnamefont
  {R.}~\bibnamefont {Yang}}, \bibinfo {author} {\bibfnamefont {H.}~\bibnamefont
  {Wang}}, \bibinfo {author} {\bibfnamefont {X.}~\bibnamefont {Yang}}, \ and\
  \bibinfo {author} {\bibfnamefont {M.}~\bibnamefont {Li}},\ }in\ \href@noop {}
  {\emph {\bibinfo {booktitle} {Proceedings of SAP2014, LanZhou, China}}}\
  (\bibinfo {year} {2014})\ pp.\ \bibinfo {pages} {12--15}\BibitemShut
  {NoStop}%
\bibitem [{\citenamefont {Billen}\ and\ \citenamefont
  {Young}(1993)}]{poissonsuperfish}%
  \BibitemOpen
  \bibfield  {author} {\bibinfo {author} {\bibfnamefont {J.~H.}\ \bibnamefont
  {Billen}}\ and\ \bibinfo {author} {\bibfnamefont {L.~M.}\ \bibnamefont
  {Young}},\ }in\ \href@noop {} {\emph {\bibinfo {booktitle} {Particle
  Accelerator Conference, 1993., Proceedings of the 1993}}}\ (\bibinfo
  {organization} {IEEE},\ \bibinfo {year} {1993})\ pp.\ \bibinfo {pages}
  {790--792}\BibitemShut {NoStop}%
\bibitem [{\citenamefont {Fl{\"o}ttmann}(2011)}]{Floettmann2011}%
  \BibitemOpen
  \bibfield  {author} {\bibinfo {author} {\bibfnamefont {K.}~\bibnamefont
  {Fl{\"o}ttmann}},\ }\href@noop {} {\emph {\bibinfo {title} {ASTRA: A space
  charge tracking algorithm}}} (\bibinfo {year} {2011})\BibitemShut {NoStop}%
\bibitem [{poi(2013)}]{pointgrey2013}%
  \BibitemOpen
  \href@noop {} {\emph {\bibinfo {title} {Flea3 GE Technical Reference
  Manual}}},\ \bibinfo {organization} {Point Grey Research Inc} (\bibinfo
  {year} {2013})\BibitemShut {NoStop}%
\bibitem [{\citenamefont {Rao}\ and\ \citenamefont
  {Dowell}(2013)}]{rao2014engineering}%
  \BibitemOpen
  \bibfield  {author} {\bibinfo {author} {\bibfnamefont {T.}~\bibnamefont
  {Rao}}\ and\ \bibinfo {author} {\bibfnamefont {D.~H.}\ \bibnamefont
  {Dowell}},\ }\href@noop {} {\emph {\bibinfo {title} {An engineering guide to
  photoinjectors}}}\ (\bibinfo  {publisher} {CreateSpace Independent Publishing
  Platform},\ \bibinfo {year} {2013})\BibitemShut {NoStop}%
\bibitem [{\citenamefont {Clark}(1975)}]{Clark1975}%
  \BibitemOpen
  \bibfield  {author} {\bibinfo {author} {\bibfnamefont {M.~G.}\ \bibnamefont
  {Clark}},\ }\href {http://stacks.iop.org/0022-3727/8/i=5/a=013} {\bibfield
  {journal} {\bibinfo  {journal} {Journal of Physics D: Applied Physics}\
  }\textbf {\bibinfo {volume} {8}},\ \bibinfo {pages} {535} (\bibinfo {year}
  {1975})}\BibitemShut {NoStop}%
\end{thebibliography}%
\end{document}